\documentclass[12pt,letterpaper]{JHEP3}
\pdfoutput=0
\usepackage{cite}
\usepackage{epsfig}
\usepackage{subfigure}

\graphicspath{{Figures/}}

\title{Rolling Through a Vacuum}

\author{Jan Pieter van der Schaar\footnote{j.p.vanderschaar@uva.nl} and I-Sheng Yang\footnote{isheng.yang@gmail.com}\\
IOP and GRAPPA, Universiteit van Amsterdam, \\
Science Park 904, 1090 GL Amsterdam, Netherlands
}

\abstract{We clarify under what conditions slow-roll inflation can continue almost undisturbed, while briefly evolving through a (semi-classically) metastable false vacuum. Furthermore, we look at potential signatures in the primordial power spectrum that could point towards the existence of traversed metastable false vacua. Interestingly, the theoretical constraints for the existence of traversable metastable vacua imply that Planck should be able to detect the resulting features in the primordial power spectrum. In other words, if Planck does not see features this immediately implies the non-existence of metastable false vacua rolled through during the inflationary epoch.}

\begin{document}

\section{Introduction}

One pending observation in cosmology is the existence of a metastable vacuum different from our own. Up to now the possibilities discussed in the literature are limited to primordial events that took place before, and set the initial conditions of,  slow-roll inflation. For instance bubble nucleation resulting in detectable levels of negative curvature \cite{FreKle05,YamLin11,KleSch12,GutYas12} or bubble collisions leading to specific patterns in the Cosmic Microwave Background sky \cite{AguJoh09,Kle11}. As a consequence a long enough period of slow-roll inflation will erase all memory of these events and leave us with nothing to observe.

A possibility that we would like to consider here is to have a metastable vacuum {\it during} slow-roll inflation.  At first this sounds quite contradictory.  Indeed, slow-roll inflation is an attractor solution involving a slowly changing state that eventually comes to and end because the slow change can no longer be maintained.  A false vacuum by definition implies an attractor solution that is trapped in a non-changing state.  Na\"ively, one might think these two attractor solutions should be mutually exclusive.  In other words, a slow-roll inflationary trajectory  should not be able to ``pass through'' a metastable false vacuum without being trapped there.

We will show that the above intuition is incorrect.  First of all we should note that this is not new.  After all there are a large number of slow-roll inflationary models and it should not come as a surprise that some of them, in some parameter range, do contain one or more metastable false vacua.  In particular there has been a lot of work over recent years on adding specific features to simple slow-roll models and study the observable outcome \cite{CheEas06,Che10}. Although the stand-alone properties of these features in the potential are typically not of main concern and therefore less studied, in some parameter range they actually do correspond to metastable false vacua, as we will point out. We will use the simplest possible model -- slow-roll with a canonical scalar field -- to illustrate the dynamics of rolling through a metastable false vacuum. As one should expect vacuum stability requires a `big' barrier in the potential. At the same time, the barrier cannot be too `big' to disrupt the slow-roll process.  By carefully defining what turn out to be different standards of `big' in this simplest scenario, we show that it is possible to satisfy both conditions at the same time.

Our basic example allows us to derive roughly the necessary conditions for traversing a stable vacuum during slow-roll inflation. Specifically we identify under what conditions a stable vacuum only fractionally disturbs the slow-roll process by a factor of $10^{-1}\sim10^{-2}$.  The leading order instability of the false vacuum will then be through Coleman-deLuccia tunneling, suppressed by $e^{-10}\sim e^{-100}$.  Including the observational constraints, this happens to be the parameter range that is consistent with the WMAP7\cite{WMAP7,WMAP7a} results and can be probed by Planck\cite{Planck1}. In general, a feature corresponding to a traversable false vacuum will result in an oscillating pattern on the power spectrum and the bispectrum \cite{CheEas06,AdsDvo11}. Notably slow-roll disturbances smaller than a factor of $10^{-1}\sim10^{-2}$ will not only be unobservable by Planck, but at the same time result in an unstable `vacuum'. This suggests that unlike many other properties of slow-roll inflationary models, passing through a metastable false vacuum feature can actually be ruled out. On the other hand, when a feature would be detected by Planck one will be hard-pressed to uniquely identify it as due to a metastable false vacuum. For that one would need to know the details of the shape and the phase of the oscillation, which is notoriously difficult.

Our basic result automatically generalizes to multi-field inflationary models as long as the slow-roll process follows the gradient flow and the effective field along the gradient flow has a canonical kinetic term. Although this includes a significant portion of the available models, there are some notable exceptions. It excludes slow-roll models that are not driven by gradient flow \cite{DvaKac03,Yan12}, models with effective DBI kinetic terms \cite{SilTon03} and models that cannot be described by effective scalar fields \cite{MarAds12}. Even so, we would like to stress that this mechanism implies additional opportunities, not just restricted to the period of slow-roll inflation, to observe the existence of metastable false vacua within our observable cosmological history. This includes for instance the reheating process and many possibly many other interesting scenarios waiting to be further investigated.   

The structure of this paper is as follows.  In Sec.\ref{sec-old}, we discuss some existing models that can include a false vacuum, or that are in some ways relevant to such a possibility.  In Sec.\ref{sec-through}, we introduce and analyze the basic single field inflationary model and derive the necessary conditions to slowly roll through a metastable false vacuum.  In Sec.\ref{sec-obs}, we briefly discuss the observable consequences and point out that a metastable false vacuum can be consistent with the WMAP7 and recent Planck data. Even better, {\bf if Planck does not see anything, traversable metastable false vacua are ruled out.} Finally in Sec.\ref{sec-dis}, we summarize, discuss generalizations and point towards interesting future directions. 

\section{Existing Models}
\label{sec-old}

This section is effectly a short review of related slow-roll models. Readers familiar with the topic can choose to skip it and proceed to our main analysis in Section\ref{sec-through}.

\begin{figure}
\begin{center}
\includegraphics[width=8cm]{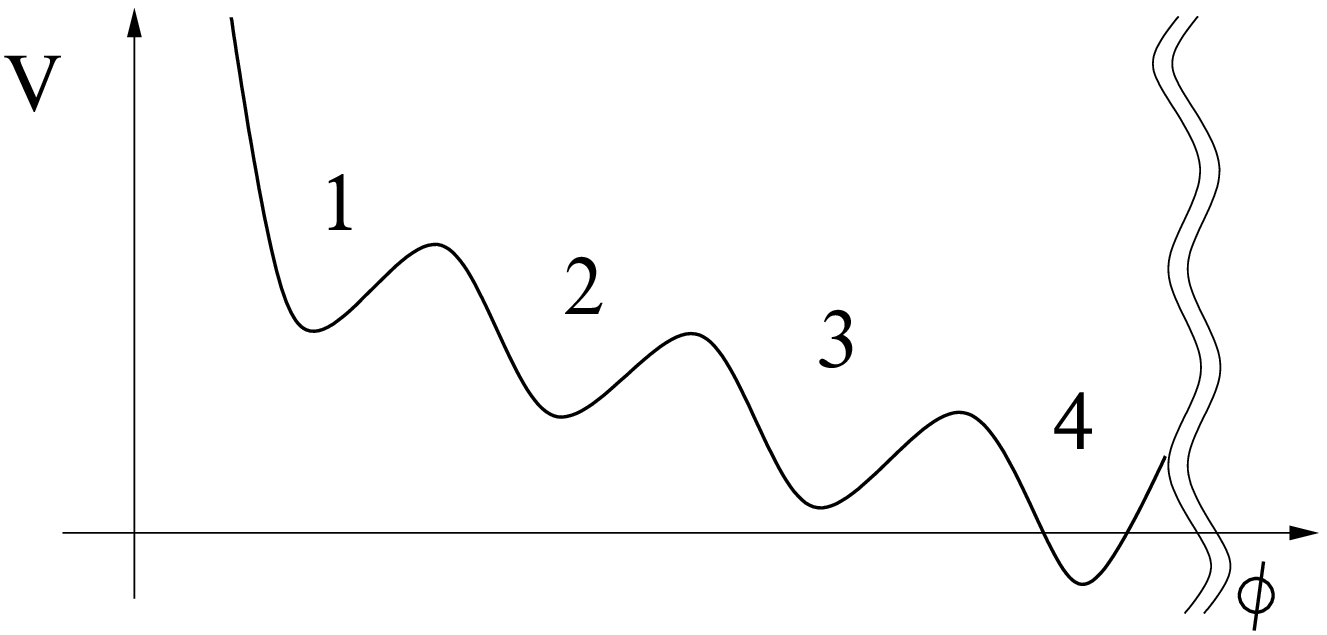}
\includegraphics[width=6cm]{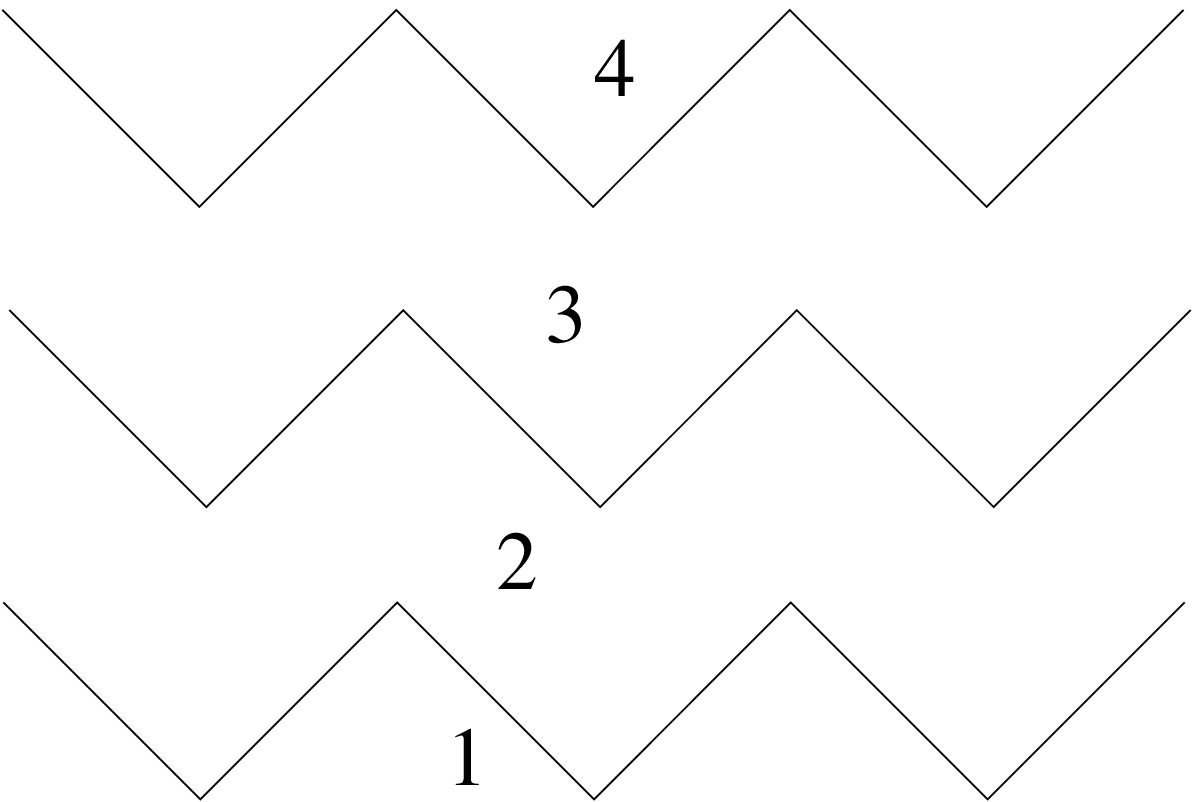}
\caption{The potential (left) and the spacetime diagram (right) of chain inflation in the original proposal.  We show only the first 4 minimum but in general there should be a lot more. The spacetime diagram depicts 3 bubbles nucleate and collide simultaneously, bringing the entire region into roughly the next minimum. Repeating these nucleations and collisions can take the field through the potential and mimic slow-roll inflation. The realistic process will not be this uniform but on large scales it is essentially the same.  
\label{fig-chain}}
\end{center}
\end{figure}

To start out let us have a look at some known models that could possibly feature the presence of metastable false vacua during inflation. An obvious set of slow-roll theories to consider are Chain-Inflation models\cite{FreSpo04}, described by (variants of) the following potential 
\begin{equation}
V(\phi) = V_0-k_c\phi - m_c^4 \cos \left( \frac{2\pi\phi}{\Delta\phi_c}\right)~.
\label{eq-chain}
\end{equation}
For $\frac{m_c^4}{\Delta\phi_c} \gg k_c$, it contains many false vacua roughly at $\phi = N\Delta\phi_c$. The essential idea is that a vacuum will rapidly tunnel to the next one by nucleating many bubbles.  These bubbles percolate and the entire spacetime transits into the next vacuum state.  Repeating this process is comparable to slow-roll evolution, as shown in Fig.\ref{fig-chain}. However, in order for the repeated percolation to take place, the typical lifetime of these false vacua should be at most one Hubble time\footnote{If the average lifetime is one Hubble time, one needs $\sim60$ minima for $\sim60$ e-foldings. However it is clear that such effective slow-roll will be quite uneven. In order to produce the observed density perturbation on needs $\sim10^4$ of repeated nucleations\cite{FelTwe06}. Later it was realized that repeated nucleation is actually not the correct dynamics when domain walls collide and the correct number of minima should be $\sim10^8$\cite{CliMoo11,KleSchun}.}. As such they do not qualify as metastable false vacua. However, there are two ways to modify this model that will allow the presence of metastable false vacua. 

\begin{figure}
\begin{center}
\includegraphics[width=8cm]{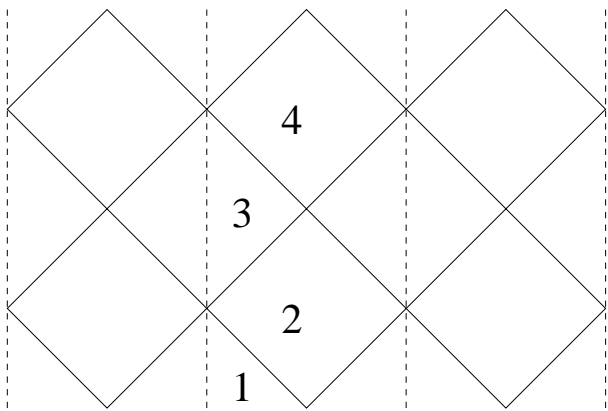}
\caption{The corrected spacetime diagram of chain inflation.  When two domain wall collides, a classical transition automatically brings the field down to the next vacuum.  Only the first generation of bubbles are necessary.  If one spatial direction is periodic, the first generation can even contain only one bubble.  One can picture such situation by periodically identifying the dashed lines.
\label{fig-chainCT}}
\end{center}
\end{figure}

The first modification comes from better understanding the classical transitions whenever domain walls collide\cite{BlaJos09,EasGib09,GibLam10,JohYan10}. For the potential given by Eq.~(\ref{eq-chain}), percolating bubbles will not just stop at the next vacuum.  The domain walls will cross each other and automatically proceed down the chain \cite{CliMoo11}, as shown in Fig.\ref{fig-chainCT}.  This implies that even though the initial false vacuum has to be short-lived to allow the first generation of bubbles to percolate, subsequently all other false vacua can have exponentially long lifetimes. Once started, the classical crossing of subsequent domain walls walls automatically results in an effective phase of slow-roll inflation.  

Recently this mechanism was further improved by introducing compact extra dimensions \cite{Bro08,DAmGob12,DAmGob12a}.  Essentially a periodic identification in Fig.\ref{fig-chainCT} will mean that an initial bubble will collide with itself in the compact dimensions. As a consequence one bubble is enough to start the entire process and all of the false vacua can have long lifetimes. Obviously these models are quite interesting, but for our purposes here they are unnecessarily contrived. For one they rely on the special mechanism of classical domain wall transitions, whereas we would like to show that passing through a metastable false vacuum does not require a special mechanism and is quite a generic possibility.

Another modification, or parameter regime, in Chain-Inflation models is to consider small $m_c$ in Eq.~(\ref{eq-chain}), such that 
\begin{equation}
\frac{2\pi m_c^4}{\Delta\phi} < k_c~.
\label{eq-mono}
\end{equation}
This makes the potential monotonically decreasing, so obviously tuning $k_c$ can lead to a standard slow-roll model with small periodic disturbances. This type of potential is in fact motivated by monodromy inflation and the periodic feature can lead to interesting signatures \cite{SilWes08}.  Most of the work in this direction assumes Eq.~(\ref{eq-mono}), but a priori that is not a necessary restriction.  If we increase $m_c$ slightly such that $V$ is no longer monotonically decreasing, it does not mean that the slow-roll evolution is suddenly significantly disturbed.  Indeed, interestingly we can keep increasing $m_c$ until the false vacua become metastable, while at the same time keeping the slow-roll process minimally disturbed.

In order to demonstrate that a metastable false vacuum does not necessarily ruin slow-roll evolution, we clearly do not have to consider a potential with many minima as in Eq.~(\ref{eq-chain}).  It can simply be understood by studying the dynamics of a slow-roll potential featuring one false vacuum. This connects nicely to studies of potentials with step features, which in some cases might correspond to a false vacuum \cite{CheEas06,AdsDvo11}.  In the next section we will show that the condition for a minimum to correspond to a metastable false vacuum is directly related to how much it disturbs the slow-roll process. As a corollary we conclude that slow-roll evolution in inflationary models with step features, or in models of monodromy inflation, can be maintained even in parameter ranges where the false vacua are metastable. 

\section{Rolling Through a Vacuum}
\label{sec-through}

\begin{figure}
\begin{center}
\includegraphics[width=8cm]{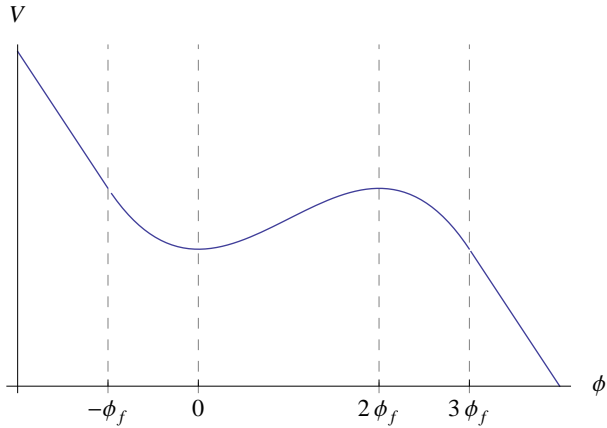}
\caption{Inserting a twisted segment in a slow-roll potential.
\label{fig-vacuum}}
\end{center}
\end{figure}

In this section we will derive the conditions for (almost) undisturbed slow-roll evolution while passing through a metastable false vacuum. To keep things as general as possible we study the simplest example with just one canonical scalar field. Let us consider the following potential, that includes a potentially metastable false vacuum in the field range between $-\phi_f<\phi<3\phi_f$, 
\begin{equation}
V(\phi) = \frac{m^2\phi^2}{2}\left(\frac{3\phi_f-\phi}{3\phi_f}\right)
+V_0~.
\end{equation}
As shown in Fig.\ref{fig-vacuum}, there is a local minimum at $\phi=0$.  The barrier is described by two parameters, its height and its width, which in our case can be identified with 
\begin{eqnarray}
{\rm barrier \ width} &\sim& \phi_f~, \\
{\rm barrier \ height} &\sim& m^2\phi_f^2~.
\end{eqnarray}
Moreover, this potential has the convenient property that at $\phi=-\phi_f$ and $\phi=3\phi_f$, the false vacuum feature connects smoothly to the unperturbed slope
\begin{equation}
-\frac{\partial V}{\partial \phi}\bigg|_{-\phi_f~{\rm or}~3\phi_f} 
=\frac{3}{2}m^2\phi_f  \equiv k~.
\label{eq-k}
\end{equation}
Clearly, this particular set-up should be well-suited to study the (slow-roll) dynamics in the presence of a false vacuum. By construction the false vacuum feature stops disturbing slow-roll when the slope of the potential returns to the original value $k$ just before it entered the feature. 

As it turns out the dynamics is most conveniently parametrized by two dimensionless parameters, $\alpha$ and $\beta$, that we will shortly define. The parameter $\alpha$ controls the stability of the false vacuum, with $\alpha>1$ corresponding to metastability. The parameter $\beta$ controls the disturbance of the slow-roll dynamics, with $\beta\gg1$ corresponding to undisturbed slow-roll evolution.

\subsection{Disturbed slow-roll evolution}

Let us assume slow-roll evolution just before the scalar field enters the false vacuum feature. Just before entering the feature we identified the slope of the potential as $k$ and demanding that the first slow-roll condition is satisfied we obtain 
\begin{equation}
\epsilon \equiv \frac{|\dot{H}|}{H^2} = M_p^2 \left( \frac{k^2}{2 V_i^2} \right) \ll1 ~,
\end{equation}
where $M_p^2 \equiv \frac{1}{8 \pi G}$ and $V_i$ is the value of the potential just before entering the false vacuum feature.

We will require that the false vacuum feature only slightly disturbs the slow-roll process. In order for this to be possible, the range in field space $\sim\phi_f$ of the barrier must be smaller than the field range corresponding to a single e-fold. Using that $k \approx -3H \dot{\phi}$ in the slow-roll limit, this means 
\begin{equation}
\phi_f \lesssim M_p^2 \left( \frac{k}{V_i} \right)~.
\end{equation}
Assuming this condition is satisfied, passing through this feature will take less than one Hubble time and the slow-roll phase should not be significantly disturbed. Correspondingly the Hubble friction is not draining a significant amount of energy and we can roughly estimate the amount of energy change caused by the false vacuum feature
\begin{equation}
\Delta \left(\frac{\dot\phi^2}{2}\right) \sim \frac{2}{3}m^2\phi_f^2~.
\end{equation}  
For this to be a small perturbation, as assumed, it should be much smaller than the typical kinetic energy during slow-roll evolution, i.e.
\begin{equation}
\frac{2}{3} m^2\phi_f^2 \ll M_p^2 \, \frac{k^2}{6 V_i}~.
\label{eq-req2}
\end{equation}
So far, this has been completely general, but now we want to make use of the fact the for the potential under consideration the slope $k$ and the field value $\phi_f$ are related by Eq.~(\ref{eq-k}). Plugging this relation into Eq.(\ref{eq-req2}) we arrive at the final condition for undisturbed slow-roll evolution in the presence of a false vacuum, introducing the dimensionless parameter $\beta$, 
\begin{equation}
\beta \equiv \frac{m}{H}\gg1~.
\label{eq-beta}
\end{equation}
One can easily check that this final result is indeed self-consistent with the starting assumption, in the sense that large $\beta$ implies that the field range $\phi_f$ is traversed in far less than a single e-fold. To be precise, defining the field range traveled in a single e-fold as $\Delta \phi_1$,  one derives that 
\begin{equation}
|\phi_f| = |2 \Delta \phi_1| \, \beta^{-2} \, .
\label{fraction-efold}
\end{equation} 
In other words the fraction of Hubble time spend moving through the false vacuum is measured by $\beta^{-2}$, which for $\beta \gg1$  is much smaller than $1$.

\begin{figure}
\begin{center}
\includegraphics[width=7cm]{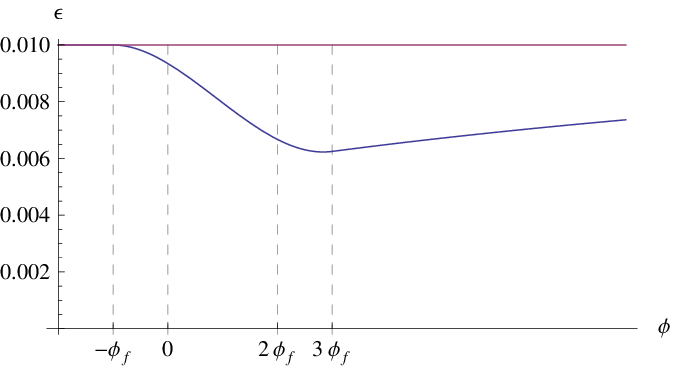}
\includegraphics[width=7cm]{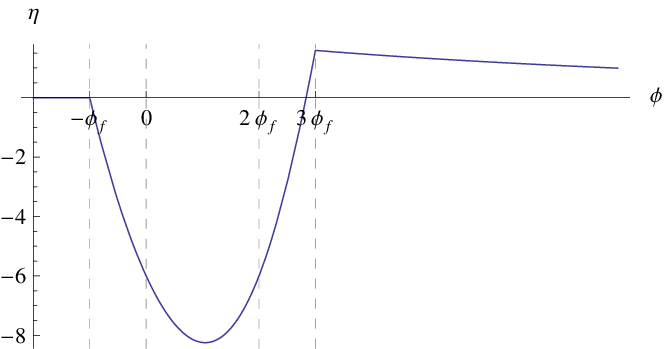}
\caption{The first slow-roll parameter $\epsilon$ (left) and the second slow-roll parameter $\eta$ (right) as functions of $\phi$ when it passes through the false vacuum.  The undisturbed values should be $\epsilon=0.01$ and $\eta=0$ in our example.
\label{fig-SRparameter}}
\end{center}
\end{figure}

A numerical solution for $\epsilon$ in Fig.\ref{fig-SRparameter} confirms our analytical estimation. Note that the second slow-roll parameter, $\eta = \dot{\epsilon}/(H\epsilon)$ is strongly disturbed, for a short duration.  That is why the spectrum of small fluctuation can be significantly affected while the overall slow-roll process is not.

\subsection{Vacuum Stability}

Now that we have derived the condition for a traversable false vacuum, let us next investigate the criteria for (meta-) stability. 
For the local minimum at $\phi=0$ to be a metastable false vacuum, we need to make sure the following two non-perturbative processes
are exponentially suppressed: 
\begin{enumerate}
\item Decay via a Hawking-Moss instanton.\footnote{Since the top of the potential barrier and the false vacuum have the same mass scale $m$, the Hawking-Moss instanton gives the same condition as any other form of thermal instability.}
\item Decay via a Coleman-deLuccia instanton.
\end{enumerate}

To find the probability for the false vacuum to decay via a Hawking-Moss instanton we compute, as usual,  the Euclidean action, giving
\begin{eqnarray}
S_{\rm HM} &=& -8\pi^2M_p^2R_{\rm HM}^2~, \\ 
3M_p^2R_{\rm HM}^{-2} &=& 3M_p^2H^2 + \frac{2}{3}m^2\phi_f^2~.
\end{eqnarray}
This result of course provides us with the exponent of the decay rate
\begin{equation}
-\ln \Gamma_{\rm HM} \sim 8\pi^2M_p^2 H^{-2} + S_{\rm HM}
\sim \frac{16\pi^2}{3}\frac{m^2\phi_f^2}{H^4}~.
\end{equation}
In order for the Hawking-Moss decay rate to be exponentially suppressed, the following condition therefore needs to be satisfied
\begin{equation}
\frac{m^2\phi_f^2}{H^4}\gg1~.
\label{eq-cond12}
\end{equation} 
We will soon see that the above condition can be considered redundant, i.e. is automatically satisfied when both the Coleman-deLuccia decay is exponentially suppressed {\it and} the false vacuum does not significantly interrupt the slow-roll evolution.  

Let us now turn to the Coleman-de Luccia decay rate. To start we first estimate the domain wall tension,
\begin{equation}
\sigma \sim m \phi_f^2~.
\end{equation}
Note however that in this case we should consider tunneling to a point on the potential with a non-zero slope instead of another minimum, implying that the thin-wall approximation is invalid. In other words, the bubble size roughly equals the thickness of the wall $m^{-1}$.  If we would nevertheless estimate the tunneling rate using the standard thin-wall formula, we obtain
\begin{equation}
-\log \Gamma_{\rm CDL} \sim \sigma m^{-3} = \frac{\phi_f^2}{m^2}~.
\label{eq-rate}
\end{equation}
A more accurate thick-wall numerical calculation\footnote{It is a numerical implementation of the overshoot-undershoot method described by Coleman\cite{CDL}. The setup is very simple and similar to \cite{BouFre06} which we will not go into details here.} of the instanton solution reveals an additional factor of $\sim300$.
\begin{equation}
-\log \Gamma_{\rm CDL} \sim 300 \frac{\phi_f^2}{m^2}~.
\end{equation} 
With the above result we now arrive at the second dimensionless parameter $\alpha$, which should satisfy the following condition to exponentially suppress Coleman-deLuccia false vacuum decay 
\begin{eqnarray}
\alpha \equiv \frac{\phi_f}{m}>1~.
\label{eq-alpha}
\end{eqnarray}

Combining conditions Eq.~(\ref{eq-alpha}) and Eq.(\ref{eq-beta}), which guarantees undisturbed slow-roll evolution, we note that the condition for exponentially suppressed Hawking-Moss decay of the false vacuum Eq.~(\ref{eq-cond12}) is automatically satisfied and therefore redundant
\begin{equation}
\frac{m^2\phi_f^2}{H^4} = \alpha^2 \, \beta^4 \gg1~.
\end{equation}
As a corollary the leading order instability of the metastable false vacuum which minimally disturbing the slow-roll process is always through Coleman-deLuccia tunneling, which is governed by Eq.~(\ref{eq-alpha}).

{\bf To summarize: in order for a false vacuum feature to be metastable and at the same time not disturbing slow-roll inflation requires that the two dimensionless parameters $\alpha \equiv \frac{\phi_f}{m}$ and $\beta\equiv \frac{m}{H}$ both be large. This corresponds to inflationary potentials with false vacua described by the following `hierarchy' of scales $\phi_f > m > H$. Such a hierarchy can easily be constructed in general, but the magnitude of primordial density perturbations does impose an additional constraint, as we will discuss next.}

\section{Observational constraints}
\label{sec-obs}

So far we have been treating $\alpha$ and $\beta$ as two free parameters, but arbitrarily tuning them will have observational consequences. This is because they implicitly control $H$ and $\epsilon$. In particular, the observed magnitude of the primordial density perturbation implies that 
\begin{equation}
\left(\frac{3}{5}\right)\left(\frac{H^2}{2\pi\dot\phi}\right)
 = \frac{9H^3}{10\pi k}
= \frac{\delta\rho}{\rho} = 10^{-5}~.
\label{eq-10-5}
\end{equation}
Combining this expression with with Eq.~(\ref{eq-k}), (\ref{eq-beta}) and (\ref{eq-alpha}), one sees that 
\begin{equation}
\alpha\beta^3 = \frac{0.6}{\pi}\frac{\rho}{\delta\rho}=\frac{6}{\pi}\times10^{4}~.
\label{eq-ab}
\end{equation}
As a consequence arbitrarily large values for both of $\alpha$ and $\beta$ are excluded. There remains some room however to satisfy both conditions, for instance for a $\beta$ between $10$ and $50$ we can still slow-roll through a metastable vacuum $\alpha\gtrsim1$. Disturbing slow-roll even less, i.e. for $\beta \gtrsim 50$, the false vacuum necessarily has to be unstable in order to satisfy the observational constraint on the primordial density perturbation. Note that the primordial density perturbation also constrains a combination of the inflationary scale $H/M_p$ and the first slow-roll parameter $\epsilon$, since $\frac{\delta\rho}{\rho} = \frac{3H}{10\pi M_p \, \sqrt{2\epsilon}}=10^{-5}$. Comparing this to Eq.~(\ref{eq-ab}) one readily sees that the parameter constraints are independent, implying that the condition to slow-roll through a metastable false vacuum does can in principle be satisfied for arbitrary inflationary scales $H/M_p$. 


We can next ask what kind of signature a traversed metastable false vacuum leaves on the CMB sky.  Fortunately there exists a large body of work on the observational signatures of isolated features on the slow-roll potential.  A particular type of feature that is very similar to a metastable false vacuum is the step feature\cite{CheEas06,AdsDvo11}. Translating the natural parameters describing a step feature (width and relative height) to the specific toy-model false vacuum studied here one finds
\begin{eqnarray}
\mathrm{width} &\sim& \phi_f~,\\
\mathrm{relative~height} &\sim& \frac{m^2\phi_f^2}{M_p^2H^2}=\frac{8\varepsilon}{\beta^2}~.
\end{eqnarray}
The original analysis applied to both negative as well as positive relative heights of the step, whereas a false vacuum can of course only be compared to a rising step feature.  As it turns out the width of the step feature in field space is not directly relevant, instead the meaningful parameter is the fraction of Hubble time the field spends in this field range, which as we saw in Eq.~(\ref{fraction-efold}) is given by $\beta^{-2}$. Looking at the above parameter identification this means that large $\beta$, besides ensuring slow-roll evolution through the false vacuum feature, should be similar to a sharp step feature. This is exactly the interesting regime studied in \cite{AdsDvo11}, so we can apply their results directly. 

According to \cite{AdsDvo11}, a sharp step feature leads to oscillating behavior in the power spectrum. The relative step height determines the amplitude and the WMAP7 constraint roughly equals \cite{WMAP7}
\begin{equation}
\frac{8\varepsilon}{\beta^2} \lesssim 0.1\varepsilon~.
\end{equation}
So we conclude that $\beta \gtrsim 10$ does not conflict with the WMAP7 observational bounds on step features. In our specific toy-model the false vacuum feature can only affect the observables by disturbing the slow-roll process. Presumably a detailed power- and bi-spectrum analysis of the recent Planck data \cite{Planck1} will improve this bound significantly.  As a consequence a false vacuum feature with $\beta \sim 10$ has a good chance to be observed. 

Most interestingly, Planck will not be able to see $\beta \gtrsim 100$ as estimated in \cite{AdsDvo11}. This is a common situation in the literature of slow-roll features---the proposed feature can still exist even if unseen by Planck.
However our case is entirely different.  Due to the constraint from $(\delta\rho/\rho)$, Eq.~(\ref{eq-ab}), larger values $\beta \gtrsim 100$ fail the metastability condition ($\alpha\gtrsim1$) for the false vacuum feature.  Therefore ``not being seen'' by Planck actually rules out the existence of a slow-roll traversed metastable false vacuum, at least in single canonical field models which are favored by recent Planck data.

If an oscillating feature in the power spectrum is observed, would it be possible to explicitly link it to a metastable false vacuum feature? Obviously this will be difficult, since we have utilized the observational bounds as derived for generic sharp step features. Most likely detection might lead to a decent estimate on the amplitude of the oscillating feature, which leaves a large degeneracy among all possible feature shapes. A metastable false vacuum is just one possibility. To really distinguish among different (sharp) features, one would need to accurately observe all details of the oscillating pattern, like the phase and the specific shape, which might be a tall order for some time to come. 

\section{Discussion}
\label{sec-dis}

In this article we explicitly determined the conditions that need to be satisfied in order for a false vacuum feature in a slow-roll potential to be 1) slow-roll traversable and 2) metastable. For this to be possible the scale $\phi_f$ corresponding to the width of the false vacuum feature has to be larger than the scale $m$ corresponding to the mass, or curvature, of the false vacuum potential, which has to be bigger than the scale of slow-roll inflation $H$, i.e. $\phi_f>m \gg H$. In general this hierarchy of scales can easily be satisfied, but the magnitude of primordial density perturbations $(\delta\rho/\rho)\sim 10^{-5}$ severely restricts the parameter space. Interestingly the remaining parameter range, expressed in terms of $\alpha$ and $\beta$, can be significantly tightened, and perhaps even excluded, by looking for oscillating signatures in the power spectrum as measured by Planck. So the existence of traversed metastable false vacua during slow-roll inflation is ruled out if Planck does not detect any oscillating features in the power spectrum. 

It would be interesting to further study the detailed observational signatures of a rolled through metastable false vacuum feature. In particular it is of interest to see whether and how a metastable false vacuum can be distinguished from other types of (sharp) features. This might in particular involve a careful analysis of higher order statistical observables of the primordial density fluctuations. The biggest degeneracy might be between a rising step (slowing down slow-roll) that could be a metastable false vacuum and a lowering step (speeding up slow-roll) that is clearly not a candidate for a metastable false vacuum. According to \cite{AdsDvo11} the difference in the observational signal is only in the initial phase of the oscillation in the power- (or bi-) spectrum. Clearly the recent Planck data set should be analyzed in detail to either rule out the existence of slow-roll traversed metastable false vacua, or in case of a feature detection, try to ascertain its properties as accurately as possible.

Let us make some final remarks regarding the `coincidence' of Planck being able to rule out the existence of traversable metastable false vacua. Our current observational capability of course should not have a fundamental physical meaning. However Eq.~(\ref{eq-ab}) provides an intriguing relation between the false vacuum stability, traversability, and the value of density perturbation. In a universe where $(\delta\rho/\rho)$ is larger, for example $10^{-1}\sim10^{-2}$, the presence of a metastable false vacuum would necessarily trap the inflaton. A trapped universe can later tunnel out. If it tunnels in the $\phi$ direction, the previous e-foldings are lost and the effective duration of inflation is reduced. If it tunnels out in other directions, most likely there will be no inflation at all\cite{FreKle05,Yan12a}. Both cases lead to a smaller chance for the standard cosmology and a bigger chance for an empty universe.

This relation might be useful in the multiverse framework\cite{Fre11}. Our value of $(\delta\rho/\rho)$ is notoriously difficult to come by as an anthropic prediction. Holding all other cosmological parameters fixed, a larger value of $(\delta\rho/\rho)$ increases the physical density of entropy/baryon/observers. This implies a na\"ive runaway problem\cite{GarVil05,BouLei09}---Our universe should have a bigger $(\delta\rho/\rho)$ if we are anthropically selected from a multiverse. Our finding points to a possible solution to that problem. If for some reason the vacuum-like features are common along a slow-roll potential, then models with larger $(\delta\rho/\rho)$ are less stable in the sense that they are prone to be trapped by false vacua and produce less observers. This implies an anthropic disadvantage for larger $(\delta\rho/\rho)$ and maybe a natural cutoff for the runaway problem.

\section*{Acknowledgemements}
We would like to thank Raphael Bousso, Ben Freivogel, Matthew Kleban, Eugene Lim, and Gary Shiu for helpful discussions. This work is supported in part by the research program of the Foundation for Fundamental Research on Matter (FOM), which is part of the Netherlands Organization for Scientific Research (NWO).

\bibliographystyle{utcaps}
\bibliography{all}

\providecommand{\href}[2]{#2}\begingroup\raggedright\begin{thebibliography}{10}

\bibitem{FreKle05}
B.~Freivogel, M.~Kleban, M.~Rodriguez~Martinez, and L.~Susskind,
  ``{Observational consequences of a landscape},'' {\em JHEP} {\bf 03} (2006)
  039,
\href{http://arxiv.org/abs/hep-th/0505232}{{\tt arXiv:hep-th/0505232}}.

\bibitem{YamLin11}
D.~Yamauchi, A.~Linde, A.~Naruko, M.~Sasaki, and T.~Tanaka, ``{Open inflation
  in the landscape},'' \href{http://dx.doi.org/10.1103/PhysRevD.84.043513}{{\em
  Phys.Rev.} {\bf D84} (2011)  043513},
\href{http://arxiv.org/abs/1105.2674}{{\tt arXiv:1105.2674 [hep-th]}}.

\bibitem{KleSch12}
M.~Kleban and M.~Schillo, ``{Spatial Curvature Falsifies Eternal Inflation},''
  \href{http://dx.doi.org/10.1088/1475-7516/2012/06/029}{{\em JCAP} {\bf 1206}
  (2012)  029},
\href{http://arxiv.org/abs/1202.5037}{{\tt arXiv:1202.5037 [astro-ph.CO]}}.

\bibitem{GutYas12}
A.~H. Guth and Y.~Nomura, ``{What can the observation of nonzero curvature tell
  us?},''
\href{http://arxiv.org/abs/1203.6876}{{\tt arXiv:1203.6876 [hep-th]}}.

\bibitem{AguJoh09}
A.~Aguirre and M.~C. Johnson, ``{A status report on the observability of cosmic
  bubble collisions},'' \href{http://arxiv.org/abs/0908.4105}{{\tt 0908.4105}}.

\bibitem{Kle11}
M.~Kleban, ``{Cosmic Bubble Collisions},''
  \href{http://dx.doi.org/10.1088/0264-9381/28/20/204008}{{\em
  Class.Quant.Grav.} {\bf 28} (2011)  204008},
\href{http://arxiv.org/abs/1107.2593}{{\tt arXiv:1107.2593 [astro-ph.CO]}}.

\bibitem{CheEas06}
X.~Chen, R.~Easther, and E.~A. Lim, ``{Large Non-Gaussianities in Single Field
  Inflation},'' \href{http://dx.doi.org/10.1088/1475-7516/2007/06/023}{{\em
  JCAP} {\bf 0706} (2007)  023},
\href{http://arxiv.org/abs/astro-ph/0611645}{{\tt arXiv:astro-ph/0611645
  [astro-ph]}}.

\bibitem{Che10}
X.~Chen, ``{Primordial Non-Gaussianities from Inflation Models},''
  \href{http://dx.doi.org/10.1155/2010/638979}{{\em Adv.Astron.} {\bf 2010}
  (2010)  638979},
\href{http://arxiv.org/abs/1002.1416}{{\tt arXiv:1002.1416 [astro-ph.CO]}}.

\bibitem{WMAP7}
E.~Komatsu {\em et al.}, ``{Seven-Year Wilkinson Microwave Anisotropy Probe
  (WMAP) Observations: Cosmological Interpretation},''
\href{http://arxiv.org/abs/1001.4538}{{\tt arXiv:1001.4538 [astro-ph.CO]}}.

\bibitem{WMAP7a}
N.~Jarosik {\em et al.}, ``{Seven-Year Wilkinson Microwave Anisotropy Probe
  (WMAP) Observations: Sky Maps, Systematic Errors, and Basic Results},''
  \href{http://dx.doi.org/10.1088/0067-0049/192/2/14}{{\em Astrophys. J.
  Suppl.} {\bf 192} (2011)  14},
\href{http://arxiv.org/abs/1001.4744}{{\tt arXiv:1001.4744 [astro-ph.CO]}}.

\bibitem{Planck1}
{\bf Planck Collaboration} Collaboration, P.~Ade {\em et al.}, ``{Planck Early
  Results. I. The Planck mission},''
  \href{http://dx.doi.org/10.1051/0004-6361/201116464}{{\em Astron.Astrophys.}
  {\bf 536} (2011)  16464},
\href{http://arxiv.org/abs/1101.2022}{{\tt arXiv:1101.2022 [astro-ph.IM]}}.

\bibitem{AdsDvo11}
P.~Adshead, C.~Dvorkin, W.~Hu, and E.~A. Lim, ``{Non-Gaussianity from Step
  Features in the Inflationary Potential},''
  \href{http://dx.doi.org/10.1103/PhysRevD.85.023531}{{\em Phys.Rev.} {\bf D85}
  (2012)  023531},
\href{http://arxiv.org/abs/1110.3050}{{\tt arXiv:1110.3050 [astro-ph.CO]}}.

\bibitem{DvaKac03}
G.~Dvali and S.~Kachru, ``{New old inflation},''
\href{http://arxiv.org/abs/hep-th/0309095}{{\tt arXiv:hep-th/0309095
  [hep-th]}}.

\bibitem{Yan12}
I.-S. Yang, ``{The Strong Multifield Slowroll Condition and Spiral
  Inflation},''
\href{http://arxiv.org/abs/1202.3388}{{\tt arXiv:1202.3388 [hep-th]}}.

\bibitem{SilTon03}
E.~Silverstein and D.~Tong, ``{Scalar speed limits and cosmology: Acceleration
  from D-cceleration},''
  \href{http://dx.doi.org/10.1103/PhysRevD.70.103505}{{\em Phys.Rev.} {\bf D70}
  (2004)  103505},
\href{http://arxiv.org/abs/hep-th/0310221}{{\tt arXiv:hep-th/0310221
  [hep-th]}}.

\bibitem{MarAds12}
E.~Martinec, P.~Adshead, and M.~Wyman, ``{Chern-Simons EM-flation},''
  \href{http://dx.doi.org/10.1007/JHEP02(2013)027}{{\em JHEP} {\bf 1302} (2013)
   027},
\href{http://arxiv.org/abs/1206.2889}{{\tt arXiv:1206.2889 [hep-th]}}.

\bibitem{FreSpo04}
K.~Freese and D.~Spolyar, ``{Chain inflation: 'Bubble bubble toil and
  trouble'},'' \href{http://dx.doi.org/10.1088/1475-7516/2005/07/007}{{\em
  JCAP} {\bf 0507} (2005)  007},
\href{http://arxiv.org/abs/hep-ph/0412145}{{\tt arXiv:hep-ph/0412145
  [hep-ph]}}.

\bibitem{FelTwe06}
B.~Feldstein and B.~Tweedie, ``{Density Perturbations in Chain Inflation},''
  \href{http://dx.doi.org/10.1088/1475-7516/2007/04/020}{{\em JCAP} {\bf 0704}
  (2007)  020},
\href{http://arxiv.org/abs/hep-ph/0611286}{{\tt arXiv:hep-ph/0611286
  [hep-ph]}}.

\bibitem{CliMoo11}
J.~M. Cline, G.~D. Moore, and Y.~Wang, ``{Chain Inflation Reconsidered},''
  \href{http://dx.doi.org/10.1088/1475-7516/2011/08/032}{{\em JCAP} {\bf 1108}
  (2011)  032},
\href{http://arxiv.org/abs/1106.2188}{{\tt arXiv:1106.2188 [hep-th]}}.

\bibitem{KleSchun}
M.~Kleban and M.~Schillo, ``{private communication},''.

\bibitem{BlaJos09}
J.~J. Blanco-Pillado, D.~Schwartz-Perlov, and A.~Vilenkin, ``{Quantum Tunneling
  in Flux Compactifications},''
\href{http://arxiv.org/abs/0904.3106}{{\tt arXiv:0904.3106 [hep-th]}}.

\bibitem{EasGib09}
R.~Easther, J.~Giblin, John~T., L.~Hui, and E.~A. Lim, ``{A New Mechanism for
  Bubble Nucleation: Classical Transitions},''
\href{http://arxiv.org/abs/0907.3234}{{\tt arXiv:0907.3234 [hep-th]}}.

\bibitem{GibLam10}
J.~T. Giblin, Jr, L.~Hui, E.~A. Lim, and I.-S. Yang, ``{How to Run Through
  Walls: Dynamics of Bubble and Soliton Collisions},''
  \href{http://dx.doi.org/10.1103/PhysRevD.82.045019}{{\em Phys. Rev.} {\bf
  D82} (2010)  045019},
\href{http://arxiv.org/abs/1005.3493}{{\tt arXiv:1005.3493 [hep-th]}}.

\bibitem{JohYan10}
M.~C. Johnson and I.-S. Yang, ``{Escaping the crunch: gravitational effects in
  classical transitions},''
  \href{http://dx.doi.org/10.1103/PhysRevD.82.065023}{{\em Phys. Rev.} {\bf
  D82} (2010)  065023},
\href{http://arxiv.org/abs/1005.3506}{{\tt arXiv:1005.3506 [hep-th]}}.

\bibitem{Bro08}
A.~R. Brown, ``{Boom and Bust Inflation: a Graceful Exit via Compact Extra
  Dimensions},'' \href{http://dx.doi.org/10.1103/PhysRevLett.101.221302}{{\em
  Phys. Rev. Lett.} {\bf 101} (2008)  221302},
\href{http://arxiv.org/abs/0807.0457}{{\tt arXiv:0807.0457 [hep-th]}}.

\bibitem{DAmGob12}
G.~D'Amico, R.~Gobbetti, M.~Schillo, and M.~Kleban, ``{Inflation from Flux
  Cascades},''
\href{http://arxiv.org/abs/1211.3416}{{\tt arXiv:1211.3416 [hep-th]}}.

\bibitem{DAmGob12a}
G.~D'Amico, R.~Gobbetti, M.~Kleban, and M.~Schillo, ``{Unwinding Inflation},''
  \href{http://dx.doi.org/10.1088/1475-7516/2013/03/004}{{\em JCAP} {\bf 1303}
  (2013)  004},
\href{http://arxiv.org/abs/1211.4589}{{\tt arXiv:1211.4589 [hep-th]}}.

\bibitem{SilWes08}
E.~Silverstein and A.~Westphal, ``{Monodromy in the CMB: Gravity Waves and
  String Inflation},'' \href{http://dx.doi.org/10.1103/PhysRevD.78.106003}{{\em
  Phys.Rev.} {\bf D78} (2008)  106003},
\href{http://arxiv.org/abs/0803.3085}{{\tt arXiv:0803.3085 [hep-th]}}.

\bibitem{CDL}
S.~Coleman and F.~D. Luccia, ``Gravitational effects on and of vacuum decay,''
  {\em Phys. Rev. D} {\bf 21} (1980)  3305--3315.

\bibitem{BouFre06}
R.~Bousso, B.~Freivogel, and M.~Lippert, ``Probabilities in the landscape: The
  decay of nearly flat space,'' {\em Phys. Rev.} {\bf D74} (2006)  046008,
\href{http://arxiv.org/abs/hep-th/0603105}{{\tt hep-th/0603105}}.

\bibitem{Yan12a}
I.-S. Yang, ``{Probability of Slowroll Inflation in the Multiverse},''
  \href{http://dx.doi.org/10.1103/PhysRevD.86.103537}{{\em Phys.Rev.} {\bf D86}
  (2012)  103537},
\href{http://arxiv.org/abs/1208.3821}{{\tt arXiv:1208.3821 [hep-th]}}.

\bibitem{Fre11}
B.~Freivogel, ``{Making predictions in the multiverse},''
  \href{http://dx.doi.org/10.1088/0264-9381/28/20/204007}{{\em
  Class.Quant.Grav.} {\bf 28} (2011)  204007},
\href{http://arxiv.org/abs/1105.0244}{{\tt arXiv:1105.0244 [hep-th]}}.

\bibitem{GarVil05}
J.~Garriga and A.~Vilenkin, ``Anthropic prediction for {L}ambda and the {Q}
  catastrophe,'' {\em Prog. Theor. Phys. Suppl.} {\bf 163} (2006)  245--257,
\href{http://arxiv.org/abs/hep-th/0508005}{{\tt hep-th/0508005}}.

\bibitem{BouLei09}
R.~Bousso and S.~Leichenauer, ``{Predictions from Star Formation in the
  Multiverse},'' \href{http://dx.doi.org/10.1103/PhysRevD.81.063524}{{\em
  Phys.Rev.} {\bf D81} (2010)  063524},
\href{http://arxiv.org/abs/0907.4917}{{\tt arXiv:0907.4917 [hep-th]}}.

\end{thebibliography}\endgroup

\end{document}